\documentstyle[11pt,dunk2001_asp,twoside,epsf]{article}
\def\edcomment#1{\iffalse\marginpar{\raggedright\sl#1\/}\else\relax\fi}
\reversemarginpar

\begin{document}
\title{Into the future with GAIA} 
\author{Tim de Zeeuw}
\affil{Leiden Observatory, The Netherlands}

\begin{abstract}
The GAIA space observatory was recently approved as Cornerstone 6 of
ESA's science program, to be launched no later than mid-2012. It will
provide a stereoscopic and kinematic census of about $10^9$ stars
throughout our Galaxy (and into the Local Group) complete to
$V=20$~mag, amounting to about 1\% of the Galactic stellar
population. Combined with astrophysical information for each star,
provided by on-board multi-colour photometry and (limited)
spectroscopy, the positional and radial velocity measurements will
have the precision necessary to quantify the early formation, and
subsequent dynamical, chemical and star formation evolution of our
Galaxy: GAIA will establish when the stars in our Galaxy formed, when
and how the Galaxy was assembled, and how the dark matter is
distributed. The GAIA data will also allow detection and orbital
classification of $\approx 5\times 10^4$ extra-Solar planetary
systems, provide a comprehensive survey of $10^5-10^6$ minor bodies in
our Solar System, of galaxies in the nearby Universe, of some $5\times
10^5$ quasars, and will test general relativity and cosmology.
\end{abstract}

\section{Structure and Evolution of the Milky Way} 

The final sentences of Ken Freeman's influential 1987 review of `The
Galactic Spheroid and Old Disk' read: {\it In this review, I have
emphasized studies attempting to elucidate the present dynamical state
of the Galaxy. This knowledge is needed before we can hope to proceed
with any confidence to the next step: understanding the chain of
events that occurred during the formation of the Galaxy}. This next
step is the main scientific goal of GAIA\footnote{The acronym has many
interpretations, with {\sl Galactic Astrophysics through Imaging and
Astrometry} and {\sl Great Advance In Astrophysics} amongst them.}, an
astrophysics mission selected as Cornerstone~6 in the ESA science
program (Perryman et al.\ 2001). A brief look into this future
opportunity for Milky Way research is therefore particularly
appropriate at this symposium in honor of Ken.

The Milky Way contains a complex mix of stars, planets, interstellar
gas and dust, and dark matter. These components are distributed in age
(reflecting their birth rate), in space (reflecting their birth places
and subsequent motions), on orbits (determined by the gravitational
force generated by their own mass), and in chemical element abundances
(determined by the past history of star formation and gas
accretion). To understand the formation, structure and evolution of
our Galaxy therefore requires three complementary observational
approaches: (i) a census of the contents of a large and representative
part of the Galaxy; (ii) quantification of the present spatial
structure, from distances; (iii) determination of the
three-dimensional motions. This can be obtained from complementary
astrometry, photometry, and radial velocities.

Astrometric measurements uniquely provide model-independent distances
(trigonometic parallaxes) and transverse kinematics (proper
motions). Multi-color photometry, with appropriate astrometric and
astrophysical calibration, provides a measure of extinction, and
hence, combined with astrometry, allows derivation of intrinsic
luminosities, spatial distribution functions, and stellar chemical
abundance and age information. Radial velocities complete the
kinematic triad, allowing determination of gravitational forces,
stellar orbits, and the distribution of invisible mass. Astrometry and
limited photometry from HIPPARCOS, supplemented with ground-based
radial velocities, provided this information for $\sim$$10^5$ stars in
one small part of the Milky Way, the immediate Solar neighborhood.
GAIA will provide a representative census of the entire Galaxy in one
fell swoop, by repeatedly measuring the positions and multi-color
brightness of {\it all} $10^9$ objects to $V=20$~mag, and radial
velocities to $V=17$~mag (Table~1). The astrometric accuracies will be
at the micro-arcsec ($\mu$as) level, and will provide distances to
better than 10\% well beyond the Galactic Center. Individual stellar
motions will be measured even in M31. On-board detection will ensure
that variable stars, supernovae, transient sources, micro-lensed
events, and minor planets will all be observed and catalogued to
$V=20$~mag.

\begin{table}[t]
\begin{center}
\caption{GAIA compared with HIPPARCOS} 
\vspace{5pt}
    \begin{tabular}{lll}
\hline && \\[-10pt]
\hline && \\[-5pt]
                   &HIPPARCOS $\quad$        &GAIA  \\
\noalign{\smallskip}
\hline && \\[-5pt]
Magnitude limit    &12                &20--21 mag \\
Completeness       &7.3--9.0          &$\sim$20 mag \\
Bright limit       &$\sim$0           &$\sim$3--7 mag \\
Number of objects  &$1.2\times 10^5$  &$2.6\times 10^7$ to $V=15$ \\
                   &                  &$2.5\times 10^8$ to $V=18$ \\
                   &                  &$1.1\times 10^9$ to $V=20$ \\
Effective distance limit &1 kpc       &1 Mpc \\
Quasars            &none              &$\sim 5\times 10^5$ \\
Galaxies           &none              &$10^6 - 10^7$ \\
Accuracy           &1 mas             &4 $\mu$as at $V=10$ \\
                   &                  &10 $\mu$as at $V=15$ \\
                   &                  &160 $\mu$as at $V=20$ \\
Broad band         &$B$ and $V$       &4--color to $V=20$ \\
Medium band        &none              &11--color to $V=20$ \\
Radial velocity    &none              &1--10 km/s to $V=16-17$ \\
Observing program  &Pre-selected      &On-board and unbiased \\[3pt]
\hline 
\end{tabular}
\label{tab:gaia-hipp}
\end{center}
\vspace{-20pt}
\end{table}

\section{Science with GAIA}

The range of scientific topics which will be addressed by the GAIA
data is vast, covering much of modern astrophysics, as well as Solar
system studies and fundamental physics. A full description can be
found in the ESA Concept and Technology Study, available on the GAIA
website {\tt http://astro.estec.esa.nl/GAIA}. Documents there contain
references to the original work briefly summarized here, as well as
details of the many other exciting scientific projects which GAIA will
address, but which space precludes discussion of here.

\subsection{The stellar halo of the Milky Way}

The ESA Concept and Technology study identified many areas of Milky
Way research where GAIA will make decisive contributions, including
the structure of the Galactic bulge, the determination of the
distribution of dark matter, the delineation of spiral structure, the
internal dynamical structure of star forming regions and globular
clusters, retracing the paths of run-away stars, etc. Here we focus on
one such area: the key role of the stellar halo in distinguishing
among competing galaxy formation scenarios. The recent review by
Freeman \& Bland-Hawthorn (2002) covers this topic in great depth.

The stellar halo contains only a small fraction of the total luminous
mass of the Galaxy, but the kinematics and abundances of halo stars,
globular clusters, and the dwarf satellites contain imprints of the
formation of the entire Milky Way. The most metal-deficient stars,
with [Fe/H]~$< -3.5$, represent a powerful tool to understand
primordial abundances and the nature of the objects which produced the
first heavy elements. The classical picture of inner monolithic
collapse, combined with later accretion in the outer Galaxy, predicts
a smooth distribution both in configuration and velocity space for our
Solar neighborhood. The currently popular theories of hierarchical
formation of structure propose that big galaxies are formed by mergers
and accretion of smaller building blocks. These events leave
signatures in the phase-space distribution of the stars that once
formed those systems but are now part of the stellar halo.

In the hierarchical scenario, the current spatial distribution of
stars in the inner halo should be fairly uniform, whereas strong
clumping is expected in velocity space. This clumping reveals itself
in the form of a large number of moving groups (several hundred in a
1~kpc$^3$ volume centered on the Sun, if the whole stellar halo were
built in this way) each having very small velocity dispersion. The
required velocity accuracies to detect individual halo streams are
less than a few~km~s$^{-1}$, requiring measurement precision of order
$\mu$as. A good way to find such streams is to use the space of
adiabatic invariants.  Here clumping should be stronger since all
stars originating from the same progenitor have very similar integrals
of motion, resulting in a superposition of the corresponding streams.
In this way, Helmi et al.\ (1999) found evidence for one halo stream
in the HIPPARCOS data, which must have resulted from an accretion
about 12 Gyr ago. The future astrometric missions DIVA and FAME will
find some additional streams, but it will take the GAIA accuracy,
limiting magnitude, and radial velocities to reconstruct the entire
set of merged satellites (Helmi \& de Zeeuw 2000).

\subsection{Stellar Astrophysics} 
 
GAIA will provide distances to an unprecedented 0.1\% accuracy for
$7\times 10^5$ stars out to a few hundred pc, and to 1\% accuracy for
a staggering $2\times 10^7$ stars up to a few kpc. Distances to better
than 10\% will reach beyond 10~kpc, and will cover a significant
fraction of our Galaxy, including the Galactic Center, spiral arms,
the halo, and the bulge, and---for the brightest stars---to the
nearest satellites.  With the parallel determination of
extinction/reddening and metallicities by means of multi-band
photometry and spectroscopy, this will provide an extended basis for
reading {\it in situ\/} stellar and galactic evolution.  All parts of
the Hertzsprung--Russell diagram will be comprehensively calibrated,
from pre-main sequence stars to white dwarfs and all transient phases;
all possible masses, from brown dwarfs to the most massive O~stars;
all types of variable stars; all possible types of binary systems down
to brown dwarf and planetary systems; and all standard distance
indicators (pulsating stars, cluster sequences, supergiants, central
stars of planetary nebulae, etc.). This extensive amount of accurate
data will stimulate a revolution in the exploration of stellar and
Galactic formation and evolution, and the determination of the cosmic
distance scale.

The GAIA large-scale photometric survey will have significant
intrinsic scientific value for stellar astrophysics, providing
$2\times 10^7$ variable stars of nearly all types, including detached
eclipsing binaries, contact or semi-contact binaries, and pulsating
stars.  The pulsating stars include key distance calibrators such as
Cepheids and RR~Lyrae stars and long-period variables. Existing
samples are incomplete already at magnitudes as bright as
$V\sim10$~mag. The GAIA samples to $V=20$ will be sufficiently large
($\sim 10^4$ Cepheids, and $\sim 10^5$ RR Lyrae's) to calibrate
period-luminosity relationships across a wide range of stellar
parameters, including metallicity.  A systematic variability search in
the entire GAIA survey will also reveal stars in short-lived but key
stages of stellar evolution, such as the helium core flash and the
helium shell thermal pulses and flashes. Prompt processing will
identify many targets for follow-up ground-based studies.

\subsection{The star formation history of the Galaxy}

A central element of the GAIA mission is the determination of the
evolution of the star formation rate, and the cumulative numbers of
stars formed, of the bulge, inner disk, Solar neighborhood, outer
disk and halo of our Galaxy. This information, together with the
kinematic information from GAIA, and complementary chemical abundance
information, again primarily from GAIA, provides the full evolutionary
history of the Galaxy.

\subsection{Binaries and extrasolar planets}

Many of the $10^9$ stars to be observed by GAIA are not single. A key
constraint on double and multiple star formation is the distribution
of mass-ratios $q$. For wide pairs ($>0\farcs5$) this follows from the
distribution of magnitude differences $\Delta m$.  GAIA will provide a
photometric determination of the $q$-distribution down to $q\sim0.1$,
covering the expected maximum around $q\sim0.2$.  Furthermore, the
large numbers of astrometric orbits will allow derivation of the
important statistics of the very smallest (brown dwarf) masses as well
as the detailed distribution of orbital eccentricities.

GAIA is very sensitive to non-linear proper motions. A large fraction
of all astrometric binaries with periods from 0.03--30 years will be
recognized by their poor fit to a standard single-star model. Most
will be unresolved, with very unequal mass-ratios and/or magnitudes,
but in many cases a photocenter orbit can be determined. For this
period range, the absolute and relative binary frequency can be
established, with the important possibility of exploring variations
with age and place of formation in the Galaxy.  Some $10^7$ binaries
closer than 250~pc will be detected, with much larger numbers still
detectable out to 1~kpc and beyond. This binary census will also cover
essentially the entire (period, $\Delta m$) diagram: the classical
`area of ignorance' between the short-period spectroscopic binaries
and the long-period visual binaries will finally be mapped.

GAIA's potential for planet detection was assessed by simulating
observations of a homogeneous set of extra-solar planetary systems, to
establish the expected sensitivity to the presence of planets and the
potential for accurate estimation of orbital parameters, as a function
of semi-major axis, period, and eccentricity, and the distance from
the Sun.  These simulations put the number of astrometric detections
of Jupiter-mass planets somewhere between 10,000--50,000, depending on
details of the detection and orbital distribution hypotheses.
Essentially all Jupiter-mass planets within 50~pc and with periods
between 1.5--9~yr will be discovered by GAIA.  Photometric detections
of planetary transits will also be a natural product of the GAIA
photometry.

\subsection{Solar System} 
 
Solar system objects present a challenge to GAIA because of their
significant proper motions, but they promise a rich scientific
reward. The minor bodies provide a record of the conditions in the
proto-Solar nebula, and their properties therefore shed light on the
formation of planetary systems. Discovery and orbital determination of
near-Earth objects is a subject of high public interest.

GAIA will detect between $10^5$ and $10^6$ new asteroids, and obtain
precise orbits for them. They will include all near-Earth objects with
diameters larger than about 1 km, as well as Trojans of Mars and
Venus.  The mission will also provide an all-sky search for Kuiper
Belt Objects, which should produce about 300 brighter than $V=20$
mag. These are remnants of the pre-Solar nebula, and form the closest
link with disks around young stellar objects.

\subsection{Extragalactic astrophysics} 
 
GAIA will make unique contributions to extragalactic astronomy,
including the structure, dynamics and stellar populations in the
Magellanic Clouds and other Galactic satellites, and in M31 and M33,
with consequences comparable to those summarized above for the Milky
Way. The faint magnitude limit and all-sky coverage allows derivation
of the space motions of Local Group galaxies, and studies of large
numbers of supernovae, galactic nuclei, and quasars.

{\bf Local Group}.  The orbits of galaxies are a result of mildly
non-linear gravitational interactions, which link the present
positions and velocities to the cosmological initial conditions.
Non-gravitational (hydrodynamic) or strongly non-linear gravitational
interactions (collisions, mergers) are sometimes significant. It will
be possible to determine reliable three-dimensional orbits for a
significant sample of galaxies in the Local Group, in a region large
and massive enough to provide a fair probe of the mass density in the
Universe. This provides direct constraints on the initial spectrum of
perturbations in the early Universe, on the global cosmological
density parameter $\Omega$, and on the relative distributions of mass
and light on length scales up to 1~Mpc.
 
{\bf Galaxies}. GAIA will provide multi-color photometry with
$\sim$$0\farcs3$ spatial resolution for all sufficiently
high-surface-brightness galaxies. This allows statistical analysis of
the photometric structure and color distribution of the central
regions of a complete, magnitude-limited sample of many tens of
thousands of galaxies with a resolution not achievable from the
ground, and study of the large-scale structure of the local
Universe. This naturally complements available redshift surveys, and
the deeper pencil-beam studies with large telescopes.

{\bf Supernovae}. GAIA will detect all point-like objects brighter
than $V=20$~mag, so that supernovae can be detected to a modulus of
$m-M\sim39$~mag, i.e., to $z\sim0.1$.  Simulations show that GAIA will
detect over $100\,000$ supernovae of all types.  Of these, the most
useful as cosmological-scale distance indicators are the Type~Ia
supernovae, whose light curves are very accurate distance
indicators. Rapid detection of such transient sources will allow
detailed ground-based determination of lightcurves and redshifts.
 
{\bf Quasars}. The astrometry to $V=20$~mag will provide a census of
$\sim500\,000$ quasars.  The mean surface density of $\sim 25\,{\rm
deg}^{-2}$ at intermediate to high Galactic latitudes will provide the
direct link between the GAIA astrometric reference system and an
inertial frame. GAIA will be sensitive to multiply-imaged quasars with
separations as small as $\sim0\farcs2$, which is the regime where most
of the lensing due to individual galaxies is expected. Photometric
variability of such systems will allow accurate measurement of the
Hubble constant, and the entire quasar sample will provide constraints
on the cosmological parameters $\Omega$ and $\Lambda_0$.

\subsection{Fundamental Physics} 
 
The dominant relativistic effect in the GAIA measurements is
gravitational light bending. Accurate measurement of the parameter
$\gamma$ of the Parametrized Post-Newtonian (PPN) formulation of
gravitational theories is of key importance in fundamental physics.
Light deflection depends on both the time-space and space-space
components of the metric tensor, and has been observed on distance
scales of $10^9-10^{21}$~m, and on mass scales from $1-10^{13}
M_\odot$, the upper ranges determined from the gravitational lensing
of quasars.  GAIA will extend these domains by two orders of magnitude
in length, and six orders of magnitude in mass. GAIA will provide a
precision of about $5\times10^{-7}$ for $\gamma$, based on multiple
observations of $\sim 10^7$ stars with $V<13$~mag at wide angles from
the Sun, with individual measurement accuracies better than
$10~\mu$as.  This accuracy is close to the values predicted by
theories that assume the Universe started with a strong scalar
component, which then relaxed to the general relativistic value with
time.
 
Other possibilities include determination of the solar oblateness,
from analysis of suitable asteroid orbits, and limiting any
gravitational wave backgrounds, from determinations of coherent jitter
in the quasar reference frame.  Gravitational waves passing over the
telescope will cause a time-varying shift in the apparent position of
a source; i.e., the waves cause apparent proper motions which are
coherent across the whole sky. GAIA could set, in the
$10^{-12}<f<10^{-10}$~Hz band, the best upper limit on $\Omega_{\rm
gw}$.

\begin{figure}[t]
\begin{center}
\leavevmode 
\plotone{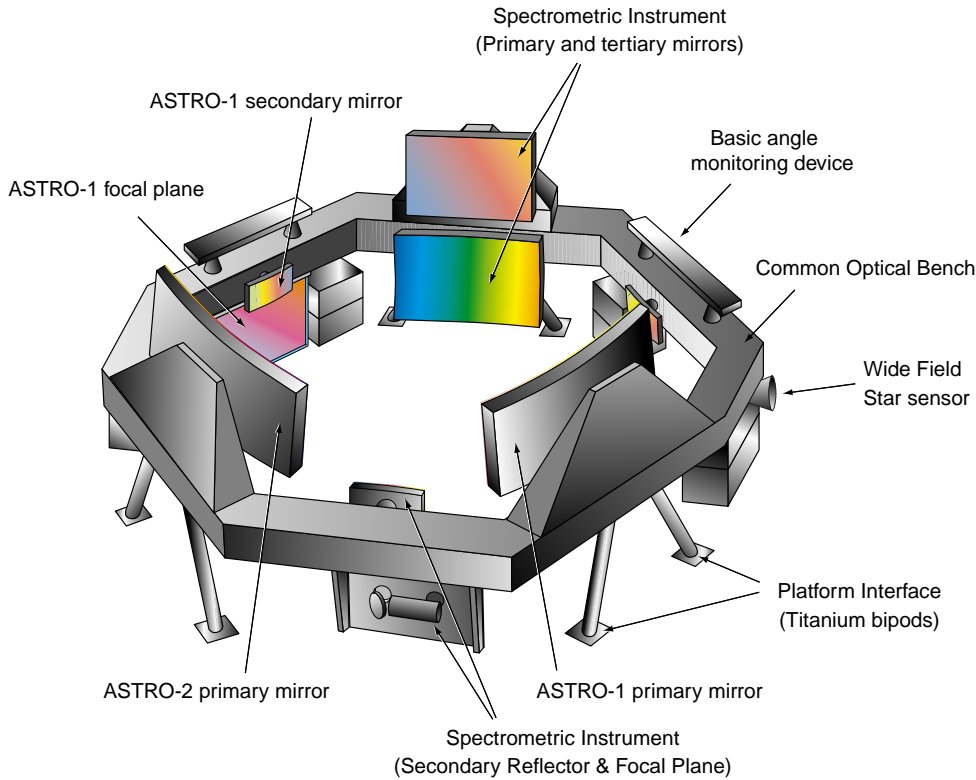}
\end{center} 
\vskip -20pt
\caption{The payload includes two identical astrometric instruments 
(labelled ASTRO-1 and ASTRO-2) separated by the 106$^\circ$ basic angle, 
as well as a spectrometric instrument (comprising a radial velocity 
measurement instrument and a medium-band photometer) which share the 
focal plane of a third viewing direction. All telescopes are 
accommodated on a common optical bench of the same material, and a 
basic angle monitoring device tracks any variations in the relative 
viewing directions of the astrometric fields. }
\end{figure}

\section{The payload} 

The proposed GAIA design has arisen from requirements on astrometric
precision (10~$\mu$as at 15~mag), completeness to $V=20$~mag, the
acquisition of radial velocities, the provision of accurate
multi-color photometry for astrophysical diagnostics, and the need
for on-board object detection. The result is a continuously scanning
spacecraft, accurately measuring one-dimensional coordinates along
great circles, and in two simultaneous fields of view, separated by a
well-defined and well-known `basic' angle. These one-dimensional
coordinates are then converted into the astrometric parameters in a
global data analysis which will provide distances and proper motions,
as well as information on double and multiple systems, photometry,
variability, metric, planetary systems, etc. The payload is based on a
large CCD focal plane assembly, with passive thermal control, and a
natural short-term (3~hour) instrument stability due to a sunshield,
the selected orbit, and a robust payload design (Figure 1).
 
The telescopes are of moderate size, with no specific design or
manufacturing complexity. The system fits within a dual-launch Ariane
5 configuration, without deployment of any payload elements. A
`Lissajous' orbit at the outer Lagrange point L2 is the preferred
operational orbit, from where an average of 1~Mbit of data per second
is returned to a single ground station throughout the planned
five-year mission. The 10~$\mu$as accuracy target has been shown to be
realistic through a comprehensive accuracy assessment program; this
remarkable accuracy is possible partly by virtue of the (unusual)
instrumental self-calibration achieved through the data analysis
on-ground. This ensures that final accuracies essentially reflect the
photon noise limit for localisation accuracy: this demanding challenge
was proven deliverable by HIPPARCOS. Figure~2 illustrates the expected
accuracy as a function of magnitude (Perryman et al.\ 2001).

\begin{figure}[t]
\begin{center}
\leavevmode
\plotone{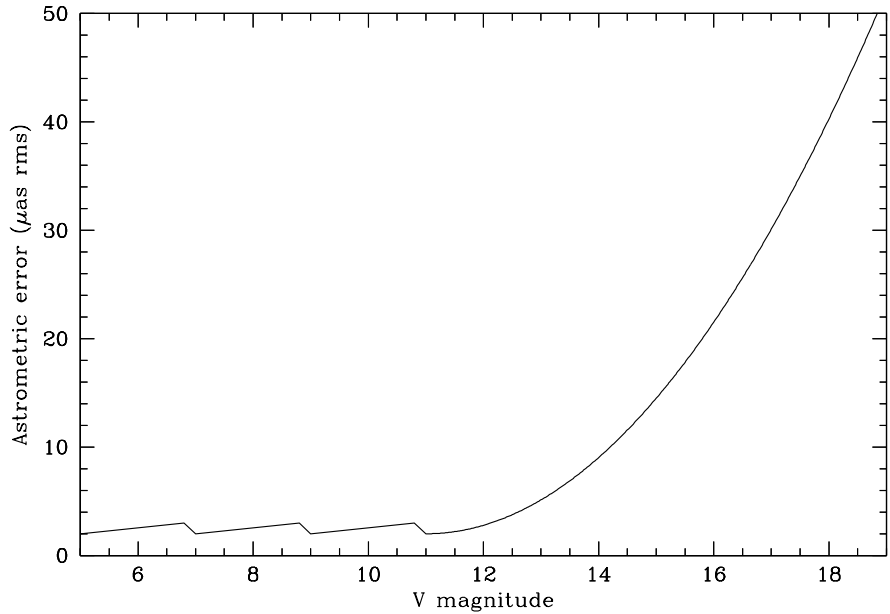}
\end{center}
\vskip -10pt
\caption{Expected GAIA accuracy versus magnitude (G2V~star).}
\end{figure}

\section{Data analysis}

The total amount of (compressed) science data generated in the course
of a five-year mission is about $2\times 10^{13}$~bytes (20~TB).  Most
of this consists of CCD raw or binned pixel values with associated
identification tags.  The data analysis aims to `explain' these values
in terms of astronomical objects and their characteristics by
iteratively adjusting the object, attitude and instrument models until
a satisfactory agreement is found between predicted and observed data.
This will require expert knowledge from several different fields of
astronomy, mathematics and computer science to be merged in a single
efficient system, including 
\begin{itemize}
\itemsep=-4truept
\item physical modeling of the observations in terms of detectors,
optics, satellite attitude and the astrometric and photometric
characteristics of the objects, including a fully general-relativistic
treatment consistent to the 1~$\mu$as level;
\item accurate geometric and photometric calibration of the
      instruments, including the celestial orientation (attitude) of the
      instrument axes;
\item efficient procedures for generating and maintaining software,
      and for the management, processing and dissemination of data.
\end{itemize} 

The ESA Concept and Technology Study considered neural network
techniques and object-orientated data structures, and included a
detailed assessment of the storage, computational processing and
algorithmic demands of the resulting satellite data stream. The
simulations have supplied confidence that, while challenging,
efficient data reduction is feasible, assuming conservative
projections of recent developments in storage devices and
computational capabilities.

\section{The GAIA Observatory} 

The GAIA study demonstrated that star selection can be effectively
undertaken autonomously on-board, which has the fundamental advantage
that GAIA science targets will be complete and unbiased. It also
eliminates the need for a complex and costly pre-launch program of
observation definition: science operations associated with the mission
will also be simplified correspondingly.

Every one of the 10$^9$ GAIA targets will be observed typically 100
times, each time in a complete set of photometric filters, and a large
fraction also with a radial velocity spectrograph. The spatial
resolution exceeds that available in ground-based surveys.  Source
detection happens on-board at each focal-plane transit, so that
variable and transient sources are detected. All these complementary
datasets, in addition to the superb positional and kinematic accuracy
which is derivable from their sum, make GAIA an optimal observatory
mission: every observable source will be observed every time it
crosses the focal plane.\looseness=-2
 
These data allow studies from asteroids to distant supernovae, from
planets to galaxies, and naturally interest almost the entire
astronomical community. For this reason, GAIA will be an open
observatory mission, directly making available its rich scientific
resource to the sponsoring communities. The scale of the GAIA data is
such that many analyses can be undertaken during operations, while
others will await final data reduction. The GAIA observatory will
provide exciting scientific data to a very wide community, beginning
with the first photometric observations, and rapidly increasing until
the fully reduced GAIA data become available. The resulting analyses
will provide a vast scientific legacy.

It is useful to compare GAIA with other astrometric missions. NASA's
SIM, to be launched around 2009, will be an interferometer, ideal for
precise measurements of a small number of carefully pre-selected
targets of specific scientific interest, focussed towards searches for
low-mass planets around a few nearby stars, calibration of the
distance scale, and detailed studies of known micro-lensing
events. FAME, a NASA MIDEX mission, and DIVA, a small German
satellite, both to be launched in 2004, are essentially successors to
HIPPARCOS, with an extension of limiting sensitivity, sample size and
accuracy by a factor of order 100 in each. They will both
substantially improve calibration of the distance scale and the main
phases of stellar evolutionary astrophysics, and map the Solar
neighborhood to much improved precision. GAIA exceeds these two
missions in scale by a further factor of order 100, allowing study of
the entire Galaxy, and only GAIA will provide photometric and radial
velocity measurements as crucial astrophysical diagnostics.

\section{Concluding remarks} 

GAIA will make it possible to create an extremely precise
three-dimensional map of a representative sample of stars throughout
our Galaxy and beyond. In the process, by combining positional data
with complementary radial velocities, GAIA will map the stellar space
motions. Through comprehensive photometric classification, GAIA will
provide the detailed physical properties of each star observed:
characterizing their luminosity, temperature, gravity, and elemental
composition. This massive multi-parameter stellar census will provide
the basic observational data to quantify the origin, structure, and
evolutionary history of our Galaxy.  The result is also of key
significance for quantitative studies of the high-redshift Universe: a
well-studied nearby template underpins analysis of unresolved galaxies
with other facilities, and at other wavelengths.

While challenging, the entire GAIA design is within the projected
state-of-the-art: the satellite is being developed in time for launch
in 2010.  By combining current technology with the demonstrated
HIPPARCOS measurement principles, GAIA will deliver an orders of
magnitude improvement in our knowledge of our Galaxy, in terms of
accuracy, number of objects, and limiting magnitude.  With this
schedule, a complete stereoscopic map of our Galaxy will be available
within 15~years. The successful completion of this program will
characterize the structure and evolution of stars and our Galaxy in a
manner completely impossible using any other methods, and nearly
inconceivable as recently as the time of Ken's 1987 review. This is an
excellent prospect to look forward to when considering Ken's 75th
birthday celebration!

\acknowledgments It is a pleasure to thank Ken and Margaret for many
years of stimulating friendship, and to thank Gary Da Costa for his
excellent organization, choice of conference venue, and patience with
a delinquent author. This paper owes much to contributions by the GAIA
Science Advisory Group, notably Michael Perryman and Gerry
Gilmore.

\end{document}